\documentstyle[prd,aps,eqsecnum,epsf,array]{revtex}

\newcommand{\be}{\begin{equation}}
\newcommand{\ee}{\end{equation}}

\newcommand{\bqn}{\begin{eqnarray}}
\newcommand{\eqn}{\end{eqnarray}}

\begin{document}

\setlength{\topmargin}{-2ex}

\draft

\title{ \bf Dressed States Approach to Quantum Systems}

\author{ G. Flores-Hidalgo$\;$\thanks{E-mail:gflores@cbpf.br}$\;$,  
A.P.C. Malbouisson $\;$\thanks{E-mail:adolfo@cbpf.br}$\;$ }  
\vspace{0.1cm} 

\address {\it Centro Brasileiro de Pesquisas F{\'\i}sicas, Rua Dr. Xavier
Sigaud 150, Urca, Rio de Janeiro CEP 22290-180-RJ, Brazil.}
 
\maketitle

\begin{abstract}

Using the non-perturbative method of {\it dressed} states previously 
introduced in \cite{JPhysA}, we study effects  of the environment on a 
quantum mechanical system, in the case the environment is modeled by an 
ensemble of non interacting harmonic oscillators. This method allows to 
separate the whole system into the {\it dressed} mechanical system and the 
{\it dressed} environment, in terms of which an exact, non-perturbative 
approach is possible. When applied to the Brownian motion, we give explicit
non-perturbative formulas for the classical path of the particle in the 
weak and strong coupling regimes. When applied to study atomic behaviours 
in cavities, the method accounts very precisely for experimentally observed
inhibition of atomic decay in small cavities \cite{PhysLA,physics0111042}.

\vspace{0.34cm}
\noindent
PACS number(s): ~03.65.Ca, 32.80.Pj 

\end{abstract}

\section{Introduction}

Quantum mechanical systems remain stable in absence of interaction. When 
interacting with an environment they lose stability as a consequence of the 
interaction. A material body, for instance an excited atom or molecule, or 
an excited nucleon, changes of state in reason of its interaction with the 
environment, the atom-electromagnetic field coupling, in the case of an atom,
or the quark-gluon interaction for a nucleon inside a nucleus. The 
understanding of the nature of the desestabilization mechanism is important
but is in general not an easy task, due to the fact it is in a large extent 
modeled by the method, in general approximate, used to study the system.
A very complete account on the subject, in particular applied to the study
of the Brownian motion can be found in Refs.\cite{zurek,paz}. From a general
point of view, in modern physics apart from computer calculations in lattice 
field theory, the only available method to treat the physics of interacting 
bodies, except for a few special cases, is perturbation theory. The 
perturbative  solution to the problem, is obtained by means of the 
introduction of bare, non interacting fields, to which are associated bare 
quanta, the interaction being introduced order by order in powers of the 
coupling constant in the perturbative expansion for the observables. The 
perturbative method gives remarkably accurate results in Quantum 
Electrodynamics and in Weak interactions. In high energy physics, asymptotic
freedom allows to apply Quantum Chromodynamics in its perturbative form and 
very important results have been obtained in this way in the last decades. 
However, in spite of its wide applicability, there are situations where the
use of perturbation theory is not possible, as in the low energy domain of 
Quantum Chromodynamics where confinement of quarks and gluons takes place, 
or are of little usefulness, as for instance in Atomic physics, in resonant
effects associated to the coupling of atoms with strong radiofrequency 
fields. These situations have led since a long time ago to attempts to 
circumvect the limitations of perturbation theory, in particular in
situations where strong effective couplings are involved. In some non 
perturbative approachs in statistical physics and constructive field theory,
general theorems can be derived using cluster-like expansions and other 
related methods \cite{Jaffe}. In some cases, these methods allow  the 
rigorous construction of field theoretical models (see for instance 
\cite{Jarrao} and other references therein), but, in spite of the rigor and
in some cases the beauty of demonstrations, they are not of  great 
usefulness in calculations of a predictive character.

As a matter of principle, due to the non vanishing of the coupling constant, 
the idea of a bare particle associated to a bare matter field is actually an
artifact of perturbation theory and is physically meaningless. A charged 
physical particle is always coupled to the gauge field, in other words, it 
is always "dressed" by a cloud of quanta of the gauge field (photons, in the 
case of Electrodynamics). In what the Brownian motion is concerned, there 
are usually two equivalent ways of modeling the environment (the thermal
bath) to which the particle is coupled: to represent the thermal bath by a 
free field, as is done in the classical work of Ref.~\cite{zurek}, or to 
consider the thermal bath as a reservoir composed of a large number of 
non-interacting harmonic oscillators (see for instance \cite{ullersma}, 
\cite{haake}, \cite{caldeira}, \cite{shram}). In both cases, exactly the 
same type of argument given above in the case of a charged particle applies
{\it mutatis} {\it mutandis} to this system, we may speak of a "dressing"
of the Brownian particle by the ensemble of the particles in the thermal 
bath. The Brownian particle should be always "dressed" by a cloud of quanta
of the thermal bath. This should be true in general for any system in which
a material body is coupled to an environment, no matter the specific nature
of environment and interaction involved.

In what follows we use the term "particle" in a general manner, a particle 
may refer for instance to an atom coupled to a field, or to a Brownian 
particle coupled to a thermal bath, the two situations where we apply our 
formalism in this paper.

In recent publications \cite{JPhysA,PhysLA,physics0111042} a method 
({\it dressed} coordinates and {\it dressed} states) has been introduced 
that allows a  non-perturbative approach to situations of the type 
described above, provided they can be approximated by a linear coupling. 
More precisely, the method applies for all systems that can be described by
an Hamiltonian of the form, 
\begin{equation} 
H=\frac{1}{2}\left[p_{0}^{2}+\omega_{0}^{2}q_{0}^{2}+ 
\sum_{k=1}^{N}(p_{k}^{2}+\omega_{k}^{2}q_{k}^{2}\right]-q_{0}
\sum_{k=1}^{N}c_{k}q_{k},  
\label{Hamiltoniana} 
\end{equation} 
where the subscript $0$ refers to the "material body" and $k=1,2,...N$ refer
to the harmonic environment modes. A Hamiltonian of this type, describing a
linear coupling of a particle with an environment, has been used in 
\cite {paz} to study the quantum Brownian motion of the particle with the 
path-integral formalism. The limit $N\rightarrow \infty$ is understood. In 
the case of the coupled atom field system, this formalism recovers the 
experimental observation that excited states of atoms in sufficiently small
cavities are stable. It allows to give formulas for the probability of an 
atom to remain excited for an infinitely long time, provided it is placed 
in a cavity of appropriate size \cite{physics0111042}. For an emission 
frequency in the visible red, the size of such cavity is in very good 
agreement with experimental observations \cite{Haroche3,Hulet}.

We give a non-perturbative treatment to the system introducing some 
{\it dressed} coordinates that allow to divide the coupled system into two 
parts, the {\it dressed} material body and the {\it dressed} environment, 
which makes unnecessary to work directly with the concepts of bare material
body, bare environment and interaction between them. In terms of these new 
coordinates {\it dressed} states are defined, which allow a non-perturbative
approach. We investigate the behaviour of the system as a function of the 
strenght of the coupling between the particle and the bath. In particular
we give explicitly non-perturbative formulas for the decay probability and 
for the classical path of the particle in the weak and strong coupling 
regimes.
 
\section{The eigenfrequencies spectrum and the diagonalizing matrix}

We consider for a moment as in \cite{JPhysA}, the problem of a harmonic 
oscillator $q_{0}$ coupled to $N$ other oscillators. In the limit 
$N\to\infty$ we recover our original situation of the coupling 
particle-bath after redefinition of divergent quantities, in a manner 
analogous as renormalization is done in field theories. The Hamiltonian 
(\ref{Hamiltoniana}) can be turned to principal axis by means of a point 
transformation, $q_{\mu}=t_{\mu}^{r}Q_{r}$, $p_{\mu}=t_{\mu}^{r}P_{r}$,
performed by an orthonormal  matrix $T=(t_{\mu}^{r})$, $\mu=(0,k)$,
$k=1,2,...\, N$, $r=0,...N$. The subscript $0$ and $k$ refer respectively 
to the particle and the harmonic modes  of the bath and $r$ refers to the 
normal modes. The transformed Hamiltonian in principal axis reads,
\begin{equation}   
H=\frac{1}{2}\sum_{r=0}^{N}(P_{r}^{2}+\Omega_{r}^{2}Q_{r}^{2}),
\label{diagonal}
\end{equation}  
where the $\Omega_{r}$'s are the normal frequencies corresponding to the 
possible collective oscillation modes of the coupled system. The matrix 
elements $t_{\mu}^{r}$ are given by \cite{JPhysA}
\begin{equation} 
t_{k}^{r}=\frac{c_{k}}{(\omega_{k}^{2}-\Omega_{r}^{2})}t_{0}^{r}\;,\;\; 
t_{0}^{r}= \left[1+\sum_{k=1}^{N}\frac{c_{k}^{2}}{(\omega_{k}^{2}-
\Omega_{r}^{2})^{2}}\right]^{-\frac{1}{2}}
\label{tkrg1} 
\end{equation} 
with the condition,
\begin{equation} 
\omega_{0}^{2}-\Omega_{r}^{2}=\sum_{k=1}^{N}\frac{c_{k}^{2}}
{\omega_{k}^{2} -\Omega_{r}^{2}}.  
\label{Nelson1} 
\end{equation} 
We take $c_{k}=\eta (\omega_{k})^{n}$. In this case environments are
classified according to $n>1$, $n=1$, or $n<1$, respectively as 
{\it supraohmic}, {\it ohmic} or {\it subohmic}. For a subohmic environment 
the sum in Eq.~(\ref{Nelson1}) is convergent and the frequency $\omega_{0}$ 
is well defined. For ohmic and supraohmic environments the sum diverges and 
a renormalization procedure is needed. In this case, after some subtraction 
steps it can be seen that Eq.~(\ref{Nelson1}) can be rewritten in the form,
\begin{equation} 
\bar{\omega}^{2}-\Omega_{r}^{2}=\eta^{2}\sum_{k=1}^{N}\frac{\Omega^{2E[n]}}
{\omega_{k}^{2} -\Omega_{r}^{2}}.  
\label{Nelson2} 
\end{equation}
We take the constant $\eta$ as $\eta=\sqrt{2g\Delta\omega}$, $\Delta\omega$
being the interval between two neighbouring bath frequencies (supposed 
uniform) and where $g$ is some constant [with dimension of 
$(frequency)^{2-\eta}$]. In the above equation  we have defined the {\it 
renormalized} frequency $\bar{\omega}$ by, 
\be
\bar{\omega}^{2}=\omega_{0}^{2}-\delta \omega^{2}\;\;\;\;\;\delta 
\omega^{2}=\frac{\eta^{2}}{4}\sum_{k=1}^{N}\sum_{\alpha=1}^{E[n]}
\Omega^{2(\alpha-1)}\omega_{k}^{2(n-\alpha)},
\label{omegabarra1}
\ee
with $E[n]$ standing for the smallest integer containing $n$. From an 
analysis of Eq.~(\ref{Nelson1}) it can be seen that if $\omega_{0}^{2}>
\delta \omega^{2}$ Eq.~(\ref{Nelson1}) yields only positive solutions for 
$\Omega^{2}$, while if $\omega_{0}^{2}<\delta \omega^{2}$, 
Eq.~(\ref{Nelson1}) has a negative solution $\Omega_{-}^{2}$. This means that
there is a damped collective normal mode that does not allows stationary 
configurations. We will not consider this last situation. Nevertheless it 
should be remarked that in a different context, it is precisely this runaway
solution that is related to the existence of a bound state in the 
Lee-Friedrechs model. This solution is considered in Ref.~\cite{Likhoded} in
the framework of a model to describe qualitatively the existence of bound 
states in particle physics. For reasons that will become apparent later, we 
restrict ouselves to the physical situations in which the environment 
frequencies $\omega_k$ can be written in the form 
\be
\omega_k=2k\pi/L,\;\;\;\;k=1,2,...\;.
\label{discreto}
\ee
Then using the formula, 
\begin{equation}
\sum_{k=1}^{N}\frac{1}{(k^{2}-u^{2})}= \left[\frac{1}{2u^{2}}-
\frac{\pi}{u}{\rm cot}(\pi u)\right],
\label{id4}
\end{equation}
Eq.~(\ref{Nelson2}) can be written in closed form,
\begin{equation} 
\mathrm{cot}(\frac{L\Omega}{2c})=\frac{\Omega^{3}}{\pi g\Omega^{2E[n]}}+
\frac{c}{L\Omega}(1-\frac{\bar{\omega}^{2}L\Omega^{2}}{\pi 
gc\Omega^{2E[n]}}).  
\label{eigenfrequencies2} 
\end{equation}

For an ohmic environment we have $c_{k}=\eta \omega_{k}$ and $\delta 
\omega^{2}=N\eta^{2}$. Taking in Eq.~ (\ref{omegabarra1}) $\omega_{0}^{2}>N
\eta^{2}$, the {\it renormalized} oscillator frequency $\bar{\omega}$ is 
given by,  
\begin{equation} 
\bar{\omega}_{ohmic}=\sqrt{\omega_{0}^{2}-N\eta^{2}},  
\label{omegabarra} 
\end{equation}
and the eigenfrequencies spectrum for an {\it ohmic} environment is given 
by the equation, 
\begin{equation} 
\mathrm{cot}(\frac{L\Omega}{2c})=\frac{\Omega}{\pi g}+\frac{c}
{L\Omega}(1-\frac{\bar{\omega}^{2}L}{\pi gc}),  
\label{eigenfrequencies1} 
\end{equation}
The solutions of Eq.~(\ref{eigenfrequencies1}) or 
Eq.~(\ref{eigenfrequencies2}) with respect to  $\Omega$ give the spectrum of 
eigenfrequencies $\Omega_{r}$ corresponding to the collective normal modes.

The transformation matrix elements turning the material body-bath system to
principal axis is obtained, after some rather long but straightforward 
manipulations analogous as it has been done in \cite{JPhysA}. They read, 
\begin{equation} 
t_{0}^{r}=\frac{\eta \Omega_{r}}{\sqrt{(\Omega_{r}^{2}-\bar{ 
\omega}^{2})^{2}+\frac{\eta^{2}}{2}(3\Omega_{r}^{2}-\bar{\omega}^{2})+
\pi^{2}g^{2}\Omega_{r}^{2}}}\;,\;\;\;   
t_{k}^{r}=\frac{\eta\omega_{k}}{\omega_{k}^{2}-\Omega_{r}^{2}}t_{0}^{r}. 
\label{t0r2} 
\end{equation}

\section{The dressed particle in an ohmic environment}

To fix our framework and to give precise applications of our formalism, we 
study in this paper an {\it ohmic} environment. The normalized eigenstates 
of our system (eigenstates of the Hamiltonian in principal axis) can be 
written in terms of normal coordinates,    
\begin{equation} 
\left<Q|n_{0},n_{1},...\right>\equiv \phi_{n_{0}n_{1}n_{2}...}(Q,t)=
\prod_{s}\left[\sqrt{\frac{2^{n_s}}{n_s!}}H_{n_{s}}(\sqrt{\frac{ 
\Omega_{s}}{\hbar}}Q_{s})\right]
\Gamma_{0}(Q)e^{-i\sum_{s}n_{s}\Omega_{s}t}, 
\label{autofuncoes} 
\end{equation} 
where $H_{n_{s}}$ stands for the $n_{s}$-th Hermite polynomial and 
$\Gamma_{0}$ is the normalized vacuum eigenfunction.
Next we intend to divide the system into the {\it dressed} particle and the
{\it dressed} environment by means of  some conveniently chosen 
{\it dressed} coordinates, $q_0'$ and $q_j'$ associated respectively to the
{\it dressed} particle and to the {\it dressed} oscillators composing the 
environment. These coordinates will allow  a natural division of the system
into the {\rm dressed} (physically observed) particle and into the {\it 
dressed} environment. The {\rm dressed} particle will contain automatically
all the effects of the environment on it. Clearly, these dressed coordinates
should not be introduced arbitrarilly. Since our problem is linear, we will 
require a linear transformation between the normal and {\it dressed} 
coordinates. Also, we demand the physical condition of vacuum stability. We 
assume that at some given time ($t=0$) the system is described by {\it 
dressed} states, whose wavefunctions are, 
\begin{equation} 
\psi_{\kappa_{0} \kappa_{1}...}(q^{\prime})=\prod_{\mu}\left[
(2^{-\kappa_{\mu}}\kappa_{\mu}!)^{-\frac{1}{2}} H_{\kappa_{\mu}} 
(\sqrt{\frac{\bar{\omega}_{\mu}}{\hbar}} q^{\prime}_{\mu})\right]
\Gamma_{0}(q')\;,  
\label{ortovestidas1} 
\end{equation}
where $q^{\prime}_{\mu}=q^{\prime}_{0},\, q^{\prime}_{i}$, $\bar{\omega}_
{\mu}=(\bar{\omega}, \omega_{i})$ and $\Gamma_{0}$ is the invariant ground 
state eigenfunction introduced in Eq.~(\ref{autofuncoes}). Note that the 
above wavefunctions will evolve in time in a more complicated form than the
unitary evolution of the eigenstates (\ref{autofuncoes}), since these 
wavefunctions are not eigenstates of the diagonal Hamiltonian 
(\ref{Hamiltoniana}). It is precisely the non unitary evolution of these 
wavefunctions that will allow (see below) a non-perturbative study of the
radiation and dissipation processes of the particle.

In order to satisfy the physical condition of vacuum stability (invariance 
under a tranformation from normal to dressed coordinates) we remember that 
the the ground state eigenfunction of the system has the form, 
\begin{equation}
\Gamma_{0}(Q)\propto {\rm e}^{-\frac{1}{2\hbar}
\sum_{r=0}^N\Omega_r Q_r^2}\;,
\label{ad1}
\end{equation}
and we require that the ground state in terms of the {\it dressed} 
coordinates should have the form
\begin{equation}
\Gamma_{0}(q')\propto {\rm e}^{-\frac{1}{2\hbar}
\sum_{\mu=0}^N\bar{\omega}_\mu (q_\mu')^2}\;.
\label{ad2}
\end{equation}
From Eqs.~(\ref{ad1}) and (\ref{ad2}) it can be seen that the vacuum 
invariance requirement is satisfied if we define {\it dressed} coordinates 
by,
\be
\sqrt{\bar{\omega}_\mu}q_\mu'=\sum_{r=0}^Nt_\mu^{r}\sqrt{\Omega_r}Q_r\;.
\label{eq11}
\ee

As we have already mentioned above our {\it dressed} states, given by 
Eq.~(\ref{ortovestidas1}), are {\it collective} but {\it non} {\it stable}
states, linear combinations of the (stable) eigensatates (\ref{autofuncoes}) 
defined in terms of the normal modes. The coefficients of these combinations 
are given in Eq.~(\ref{eq16}) below and explicit formulas for these 
coefficients for an interesting physical situation are given in 
Eq.~(\ref{coeffN}). This gives  a complete and rigorous definition of our 
dressed states. Moreover, our dressed states have the very interesting
property of distributing the energy initially in a particular dressed state, 
among itself and all other dressed states with precise and well defined 
probability amplitudes \cite{JPhysA}. We {\it choose} these dressed states 
as physically meaningful and we test successfully this hypothesis by 
studying the radiation process by an atom in a cavity. In both cases, of a 
very large or a very small cavity, our results are in agreement with 
experimental observations. 

Having introduced {\it dressed} coordinates and {\it dressed states}, in the 
next sections we will apply these concepts to study the time evolution of 
the expectation value of the particle coordinate. 

\section{Brownian motion at zero temperature} 
 
As a first application of our formalism we consider the study of the
Brownian Motion. The Brownian particle is modeled by an harmonic oscillator
coupled to an {\it ohmic} environment, The whole system being described by 
the Hamiltonian (\ref{Hamiltoniana}). This model for the Brownian motion is 
in fact not new and  has been implemented using the path integral formalism
in for instance Refs. \cite{caldeira,shram,paz}. In this approach an 
effective action for the Brownian particle is obtained, which in general 
is very complicated and non local in time. From this effective action, an 
equation for the classical path of the Brownian particle can be derived. 
However this equation obtained from the effective action, is a very 
complicated integro-differential equation that can not be solved 
analytically. But in general terms it describes the expected damped 
behaviour of the particle.

We will approach this problem using the {\it dressed} states introduced in 
the previous section, and  we will treat in detail the case in which the 
environment is at zero temperature (what corresponds to consider the 
environment initially in its ground state). Our method will account for the
expected behaviour in a much more simpler way than the usual path integral 
approach. 

We assume as usual, that initially the Brownian particle and the environment
are decoupled and that the coupling is turned on suddenly at some given 
time, that we choose at $t=0$. Since we treat here the case in which the 
environment is at zero temperature our assumption is that the initial system 
can be described by a pure {\it dressed} state. The environment state at 
zero temperature should be described by its {\it dressed} ground state. Thus 
we can write the initial state of the system particle-environment in the 
form, 
\be
|\lambda,n_1',n_2',..;t=0\left.\right>= |\left. \lambda\right>\times 
|\left.n_1',n_2',..\right> \;.
\label{eq12}
\ee
In the above equation $|\lambda\left.\right>$ is the initial {\it dressed 
state} of the particle and $|\left.n_1',n_2',..\right>$ is the initial {\it 
dressed} state of the environment (after we will take $n_1'=n_2'=...=0$,  
corresponding to the environment at zero temperature). To proceed we recall
that the classical path in the case of the quantum harmonic oscillator is 
given by the mean value of the operator position in a coherent state. In our
formalism, we define $|\lambda\left.\right>$ as a {\it dressed} coherent 
state given  by, 
\be
|\lambda\left.\right>={\rm e}^{-|\lambda|^2/2}
\sum_{n_0'=0}^{\infty}\frac{(\lambda)^{n_0'}}
{\sqrt{n_0'!}}|n_0'\left.\right>\;,
\label{eq13}
\ee
and accordingly the classical path of the Brownian particle should be given
by the time evolution of the {\it dressed} particle position operator in the
dressed coherent state (\ref{eq12}). It is useful to examinate firstly the 
time evolution of the initial coherent dressed state as given by 
Eq.~(\ref{eq12}). Replacing Eq.~(\ref{eq13}) in Eq.~(\ref{eq12}) we obtain,
\be
|\left. \lambda, n_1',n_2',..;t=0\right>={\rm e}^{-|\lambda|^2/2}
\sum_{n_0'=0}^{\infty}\frac{\lambda^{n_0'}}
{\sqrt{n_0'!}}|\left.n_0'n_1',..\right>\;.
\label{eq14}
\ee
Now, since the eigenstates $|\left.n_0,n_1,..\right>$ form a complete basis
[stable states having eigenfunctions given by Eq.~(\ref{autofuncoes})], we 
can write Eq.~(\ref{eq12}) as
\be
|\left. \lambda, n_1',n_2',..;t=0\right>={\rm e}^{-|\lambda|^2/2}
\sum_{n_0'=0}^{\infty}\sum_{\{n_r\}}\frac{\lambda^{n_0'}}
{\sqrt{n_0'!}}T_{n_0',n_1',..}^{n_0,n_1,..}|\left.n_0,n_1,..\right>\;,
\label{eq15}
\ee
where $\{n_r\}=(n_0,n_1,n_2,..)$ and
\begin{equation}
T_{n_0',n_1',..}^{n_0,n_1,..}=\left<n_0,n_1,..|n_0',n_1',..\right>
=\int dQ\phi_{n_0,n_1,..}(Q)\psi_{n_0',n_1',..}(q')\;.
\label{eq16}
\end{equation}
Since $|\left.n_0,n_1,..\right>$ are eigenvectors of the Hamiltonian 
(\ref{diagonal}), the time evolution of Eq.~(\ref{eq15}) is given by
\be
|\left. \lambda, n_1',n_2',..;t\right>={\rm e}^{-|\lambda|^2/2}
\sum_{n_0'=0}^{\infty}\sum_{\{n_r\}}\frac{\lambda^{n_0'}}
{\sqrt{n_0'!}}T_{n_0',n_1',..}^{n_0,n_1,..}
{\rm e}^{-i\sum_{r}\Omega_{r}(n_{r}+1/2)t}|\left.n_0,n_1,..\right>\;.
\label{Eq17}
\ee
Now we can compute $q_{\lambda}'(t)$, the time dependent mean value for the 
dressed oscillator position operator, $i.e.$, the mean value of the {\it 
dressed} particle position operator taken in the {\it dressed} coherent 
state (\ref{eq12}), 
\bqn
q_{\lambda}'(t)&=&\left<\lambda, n_1',n_2',...;t|q_0'
|\lambda,n_1',n_2',...;t\right>\nonumber\\
&=&{\rm e}^{-|\lambda|^2}\sum_{n_0',m_0'}\sum_{\{n_r,m_r\}}
\frac{(\lambda^\ast)^{m_0'}}{\sqrt{m_0'!}}\frac{\lambda^{n_0'}}
{\sqrt{n_0'!}}T_{m_0',n_1',..}^{m_0,m_1,..}~\!T_{n_0',n_1',..}^{n_0,n_1,..}
{\rm e}^{-i\sum_{r}\Omega_r(n_r-m_r)t}
\left<m_0,m_1,..|q_0'|n_0,n_1,..\right>\;.
\label{eq18}
\eqn
Using Eq.~(\ref{eq11}) for $\mu=0$ and
\be
\left<m_\alpha|Q_\alpha|n_\alpha\right>=\sqrt{\frac{\hbar}{2\Omega_\alpha}}
\left(\sqrt{n_\alpha}~\!\delta_{m_\alpha,n_\alpha-1}+
\sqrt{n_\alpha+1}~\!\delta_{m_\alpha,n_\alpha+1}\right)\;.
\label{eq19}
\ee
in Eq.~(\ref{eq18}) it is easy to obtain, 
\bqn
q_{\lambda}'(t) &=&  {\rm e}^{-|\lambda|^2}\sqrt{\frac{h}{2\bar{\omega}}}
\sum_s \sum_{n_0',m_0'}\sum_{\{n_r\}}t_{0}^{s}\sqrt{n_s+1}~\!
T_{n_0',n_1',..}^{n_0,n_1,..n_\alpha..}~
\!T_{n_0',n_1',..}^{n_0,n_1,..(n_s+1)..}
\times\nonumber\\
& &~~~~~~~~~~~~~~~~~~~~~~~~~~~~~~~
\left[\frac{(\lambda^\ast)^{m_0'}}{\sqrt{m_0'!}}\frac{\lambda^{n_0'}}
{\sqrt{n_0'!}}e^{-\Omega_{s}t}+\frac{(\lambda^\ast)^{n_0'}}{\sqrt{n_0'!}}
\frac{\lambda^{m_0'}}{\sqrt{m_0'!}}e^{\Omega_{s}t}\right].
\label{eq20}
\eqn
As we have mentioned above, the situation in which the environment is at 
zero temperature corresponds to $n_1'=n_2'=...=0$. In this case from 
Eqs.~(\ref{eq16}), (\ref{ortovestidas1}), (\ref{eq11}), (\ref{autofuncoes}) 
and  with the help of the theorem \cite{Ederlyi},
\begin{equation} 
\frac{1}{n_0'!}\left[\sum_{r}(t_{\mu}^{r})^{2}\right]^{\frac{m}{2}}
H_{n_0'}\left(\frac{\sum_{r}t_{\mu}^{r}\sqrt{\frac{\Omega_{r}}{\hbar}}
Q_{r}}{\sqrt{\sum_{r}(t_{\mu}^{r})^{2}}}\right)
=\sum_{m_{0}+m_{1}+...=n_0'}\frac{(t_{\mu}^{0})^{m_{0}}
(t_{\mu}^{1})^{m_{1}}...}{m_{0}!m_{1}!...}H_{m_{0}}
(\sqrt{\frac{\Omega_{0}}{\hbar}}Q_{0})H_{m_{1}}(\sqrt{\frac{\Omega_{1}}
{\hbar} }Q_{1})...\;,  
\label{teorema Ederlyi} 
\end{equation} 
we get,  
\begin{equation} 
T_{n_0,0,0,...}^{n_{0},n_{1},n_{2},...}=\sqrt{\frac{n_0'!}
{n_{0}!n_{1}!...}} (t_{0}^{0})^{n_{0}}(t_{0}^{1})^{n_{1}}
(t_{0}^{2})^{n_{2}}\delta_{m'_{0},n_{0}+n_{1}+n_{2}+...}.  
\label{coeffN} 
\end{equation} 
Replacing Eq.~(\ref{coeffN}) in Eq.~(\ref{eq20}) we obtain after some 
straightforward calculations,
\begin{equation}
q_{\lambda}'(t) =  \sqrt{\frac{h}{2\bar{\omega}}}
 \left[\lambda f^{00}(t)+\lambda^{\ast}f^{00\ast}(t)\right],
\label{Eq20}
\end{equation}
where
\be
f^{00}(t)=\sum_{s}(t_{0}^{s})^{2}e^{-i\Omega_{s}t}\;.
\label{eq21}
\ee
From Refs. \cite{JPhysA,PhysLA,physics0111042} we recognize the function 
$f^{00}(t)$ as the probability amplitude that at time $t$ the {\it 
dressed} particle still be excited, if it was initially (at $t=0$) in the 
first excited level. We see that underlying to our dressed states formalism 
there is an unified way to study two physically different situations, the 
radiation process and the Brownian motion. In next section we shall 
investigate the radiation process of the dressed particle.

Returning to the study of the Brownian particle, we see that  to obtain an
expression for the classical path we have to perform the sum appearing in 
Eq.~(\ref{eq21}) and replace the result in Eq.~(\ref{Eq20}). If we assume, as 
it is currently done in studies of the Brownian motion, that the environment
distributes itself over the whole free space, its frequencies $\omega_{k}$ 
should have a continuous  distribution. This continuum can be realized 
simply taking the limit $L\to\infty$ in Eq.~ (\ref{discreto}). In this case 
the matrix elements $t_{0}^{r}$, given by Eq.~(\ref{t0r2}) become,
\begin{equation} 
t_{0}^{r}=\lim_{\Delta \Omega \rightarrow \infty}\frac{\sqrt{2g} \Omega 
\sqrt{\Delta\Omega}}{\sqrt{(\Omega^{2}-\bar{\omega}^{2})^{2}+\pi^{2}g^{2}
\Omega_{r}^{2}}},\label{t0r3}
\end{equation}
and the function $f^{00}(t)$ in Eq.~(\ref{eq21}) can be written in the form,
\begin{equation} 
f^{00}(t)=\int_{0}^{\infty}\frac{2g\Omega^{2}e^{-i\Omega t}\, d\Omega}
{(\Omega^{2}-\omega^{2})^{2}+\pi^{2}g^{2}\Omega^{2}}. 
\label{f00} 
\end{equation} 
Before going ahead let us define a "driving parameter" $\kappa$ by,
\be
\kappa=\sqrt{\bar{\omega}^{2}-\frac{\pi^{2}g^{2}}{4}}
\label{k}
\ee
and let us study the above integral $f^{00}(t)$ in the different cases, 
$a)\;\;\kappa^{2}>0$, $b)\;\;\kappa^{2}=0$ and $c)\;\;\kappa^{2}<0$. The 
extreme cases in $a)$ and $c)$,  $\kappa^{2}\gg 0$ or $\kappa^{2}\ll 0$ 
correspond respectively to the situations of a {\it weak} coupling between 
the particle and the environment ($g\ll \bar{\omega}$) or of a {\it strong}
coupling ($g\gg \bar{\omega}$). We get for the above situations,\\
a) $\kappa^{2}>0$
\be
f^{00}(t)=\left(1-\frac{i\pi g}{2\kappa}\right)
e^{-i\kappa t-\pi gt/2}+ 2iJ(t),
\;,
\label{eq27}
\ee
b) $\kappa^{2}=0$
\be
f^{00}(t)=\left(1-\frac{\pi gt}{2}\right)e^{-\pi g t/2}+2iJ(t)
\;,\label{eq28}
\ee
and\\ 
c) $\kappa^{2}<0$
 
\be
f^{00}(t)=\frac{1}{2}\left\{ \left(1+\frac{\pi g}{2\tilde{\omega}}\right)
e^{-(\pi g/2+|\kappa|)t}+\left(1-\frac{\pi g}{2|\kappa|}\right)
e^{-(\pi g/2-|\kappa|)t}\right\}+2iJ(t)\;,
\label{eq29}
\ee
where,
\be
J(t)=2ig\int_0^{\infty}dy\frac{y^2 e^{-yt}}{(y^2+\bar{\omega}^2)^2-
\pi^2g^2y^2}\;.
\label{J}
\end{equation}
Replacing the above equations in Eq.~(\ref{Eq20}) we obtain for the classical 
path at zero temperature the following expressions,
\be
q_\lambda'(t)=\sqrt{\frac{ \hbar\bar{n} } {2\bar{\omega}}}
\left\{\left[2\cos(\kappa t+\delta)-\frac{\pi g}
{\kappa}\sin(\kappa t+\delta)\right]e^{-\pi gt/2}
+2\sin \delta J(t)\right\}\;\;\;\;\;\;(\kappa>0)\;,
\label{eq36(1)}
\ee
\be
q_\lambda'(t)=\sqrt{\frac{ \hbar\bar{n} } {2\bar{\omega}}}
\left[2(\cos\delta)\left(1-\frac{\pi g}{2}t\right)e^{-\pi g t/2}
+2(\sin\delta) J(t)\right]\;\;\;\;\;\;(\kappa=0)\;,
\label{eq37(1)}
\ee
\be
q_\lambda'(t)=\sqrt{\frac{ \hbar\bar{n} } {2\bar{\omega}}}
\left[2(\cos\delta)\left( \cosh |\kappa|t-\frac{\pi g}{2|\kappa|}
\sinh |\kappa|t\right) e^{-\pi g t/2}+
2(\sin\delta) J(t)\right]\;\;\;\;\;\;(\kappa<0)\;.
\label{eq38(1)}
\ee
In above equations we have written $\lambda=\sqrt{\bar{n}}e^{-i\delta}$, 
with $\bar{n}$ being the mean value for the number operator in the coherent
state. Eqs.~(\ref{eq36(1)}) to (\ref{eq38(1)}) give the expected behaviour 
for the classical path of the Brownian particle. Apart from a parcel 
containing the integral $J(t)$, these equations describe the behaviour of a
damped oscillator in the three regimes corresponding to $\kappa > 0$, 
$\kappa=0$ and $\kappa < 0$, with a damping coefficient equal to $\pi g$.
The above formulas describe the exact behaviours for $\delta =0$, which 
corresponds to a real value of the coherence parameter $\lambda$. The 
integral $J(t)$ in equations (\ref{eq27}-\ref{eq29}) can be evaluated for 
large times, $t>>1/\bar{\omega}$. We obtain,
\be
J(t)\approx \frac{4g}{\bar{\omega}^4t^3}\;\\;\;\;\;\;(t\gg 
\frac{1}{\bar{\omega}})
\label{J1}
\ee
Using Eq.~(\ref{J1}) in Eq.~(\ref{Eq20}) and remarking that for very large 
times the power behaviour $\sim t^{-3}$ dominates over the exponential 
decay, we obtain identical asymptotic behaviours in the three regimes above,
\be
q_\lambda'(t)\approx \sqrt{\frac{ \hbar\bar{n} } {2\bar{\omega}}}
\frac{8 g}{\bar{\omega}^4t^3}\sin\delta\;\;\;\;\;\;(\kappa>0\;,\kappa=
0\;,\kappa<0\;;\;\;\;t\gg \frac{1}{\bar{\omega}}).
\label{eq36}
\ee

The path behaviour in the different coupling regimes can be obtained from
Eqs.~(\ref{eq36(1)}) to (\ref{eq38(1)}) and Eq.~(\ref{k}). In the strong 
coupling regime, $\kappa^{2}\ll 0$ ($g\gg \bar{\omega}$), we obtain 
\be
q_\lambda'(t)\approx \sqrt{\frac{ \hbar\bar{n} } {2\bar{\omega}}}
\cos \delta \left(\frac{2\bar{\omega}^{2}}{\pi g}\right)
e^{-\frac{\bar{\omega}^{2}t}{\pi g}}+2(\sin\delta) J(t),
\label{qstrong1}
\ee
and in the weak coupling regime, $\kappa^{2}\gg 0$ ($g\ll \bar{\omega}$) 
we obtain from Eqs.~(\ref{k}) and (\ref{eq36(1)}),
\be
q_\lambda'(t)\approx \sqrt{\frac{ \hbar\bar{n} } {2\bar{\omega}}}
\left[2\cos(\bar{\omega}t+\delta)-\frac{\pi g}
{\bar{\omega}}\sin(\bar{\omega} t+\delta)\right]
e^{-\pi gt/2}
+ 2(\sin\delta) J(t).
\label{qweak1}
\ee
We see that the behaviours are quite different in the two situations, for 
not very large values of the time t, for which the exponential decay 
dominates over the power law decay of $J(t)$: an oscillatory damped 
behaviour with time in the weak coupling regime, while in the strong 
coupling regime the expected dressed coordinate value has an exponential 
decay. Again, asymptotically $J(t)$ dominates and both behaviours are 
identical obeying a power law decay $\sim t^{-3}$.

In the next section we apply our formalism to the study of the radiation 
process.

\section{The radiation process}
In this section we study the radiation process of the {\it dressed} 
particle when it is prepared in such a way that initially it is in its 
first excited state. We shall consider two situations, the particle in 
free space and the particle confined  in a cavity of diameter $L$.

\subsection{The particle in free space}
In this case the spectrum of the frequencies $\omega_k$ has a continuous
distribution as we have seen in the last section, and the function 
$f^{00}(t)$ is given by Eqs.~(\ref{eq27})-(\ref{eq29}). Combining these 
equations with Eq.~(\ref{J1}) we obtain  for the probability that the 
dressed particle still remain in its first excited state at a time 
$t\gg 1/\bar{\omega}$, the following expressions,  
\be
|f^{00}(t)|^2=\left(1+\frac{\pi^2g^2}{4\kappa^2}\right)e^{-\pi gt}
-\frac{8g}{\bar{\omega}^4t^3}\left(\sin \kappa t+
\frac{\pi g}{2\kappa}\cos \kappa t\right)
+\frac{16g^2}{\bar{\omega}^8t^3}\;\;\;\;\;\;(\kappa>0)\;,
\label{eq33}
\ee
\be
|f^{00}(t)|^2=\left(1-\frac{\pi g}{2}t\right)^2e^{-\pi g t}+
\frac{16g^2}{\bar{\omega}^8t^3}\;\;\;\;\;\;(\kappa=0),\label{eq34}
\ee
and
\be
|f^{00}(t)|^2=\left( \cosh |\kappa| t-\frac{\pi g}{2|\kappa|}
\sinh |\kappa| t\right)^2 e^{-\pi g t}+
\frac{16g^2}{\bar{\omega}^8t^3}\;\;\;\;\;\;(\kappa<0)\;.
\label{eq35}
\ee
In the {\it weak} coupling regime $\kappa \gg 0$, we obtain from 
Eq.~(\ref{eq33}) that the probability that the particle be still excited at 
time $t\gg 1/\bar{\omega}$ if it was in the first excited level at $t=0$, 
obeys the well known exponential decay law,
\be
|f^{00}(t)|^2\approx e^{-\pi gt}\;.
\label{exp}
\ee
In the strong coupling regime, $\kappa \ll 0$, we obtain from 
Eq.~(\ref{eq35}),
\be
|f^{00}(t)|^2\approx \left(\frac{\bar{\omega}}{\pi^{2}g^{2}}\right)
e^{-\frac{2\bar{\omega}^{2}t}{\pi g}}\;.
\label{strong1}
\ee
We can see that the decay of the particle is considerably enhanced for a 
strong coupling as compared to the weak coupling. We emphasize that the 
different behaviours described in Eqs.~(\ref{exp}) and (\ref{strong1}) are 
due to the fact that in the two situations the system obey {\it different} 
decay laws and that this fact {\it can} {\it not} be inferred from 
perturbation theory. It is a consequence of the {\it dressed} states 
approach.

\subsection{Behaviour of the confined system} 
Let us now consider the {\it ohmic} system in which the particle is placed 
in the center of a cavity of diameter $L$, in the case of  a very small $L$, 
{\it i.e}, that satisfies the condition of being much smaller than the 
coherence lenght, $L<<2c/g$. We note that from a physical point of view, 
$L$ stands for either the diameter of a spherical cavity or the spacing 
between infinite paralell mirrors. To fix our framework we consider a 
spherical cavity. To obtain the eigenfrequencies spectrum, we remark that 
from a graphical analysis of Eq.~(\ref{eigenfrequencies1}) it can be seen 
that in the case of a small values of $L$, its  solutions are very near the
frequency values corresponding to the asymptots of the curve $\mathrm{cot}
(\frac{L\Omega}{2c})$, which correspond to the environment modes 
$\omega_{i}=i2\pi c/L$, except from the first eigenfrequency $\Omega_{0}$. 
The only exception is the smallest solution $\Omega_{0}$. As we take larger 
and larger solutions, they are nearer and nearer to the values corresponding
to the asymptots. For instance, for a value of $L$ of the order of $2\times 
10^{-2}m$  and $\bar{\omega}\sim 10^{10}/s$, only the lowest eigenfrequency 
$\Omega_{0}$ is significantly different from the field frequency 
corresponding to the first asymptot, all the other eigenfrequencies 
$\Omega_{k}$, $k=1,2,...$ being very close to the field modes $k2\pi c/L$. 
For higher values of $\bar{\omega}$ (and lower values of $L$) the 
differences between the eigenfrequencies and the field modes frequencies 
are still smaller. Thus to solve Eq.~(\ref{eigenfrequencies1}) for the larger
eigenfrequencies we expand the function $\mathrm{cot}(\frac{L\Omega}{2c})$ 
around the values corresponding to the asymptots. We write,
\begin{equation}
\Omega_k=\frac{2\pi c}{L}(k+\epsilon_k),~~~k=1,2,..
\label{others}
\end{equation} 
with $0<\epsilon_{k}<1$, satisfying the equation,
\begin{equation}
\mathrm{cot}(\pi \epsilon_k)=\frac{2c}{gL}(k+\epsilon_k) +
\frac{1}{(k+\epsilon_k)}
(1-\frac{\bar{\omega}^{2}L}{2\pi gc}).  
\label{eigen2} 
\end{equation}
But since for a small value of $L$ every $\epsilon_k$ is  much smaller than 
$1$, Eq.~({\ref{eigen2}) can be linearized in  $\epsilon_k$, giving, 
\begin{equation}
\epsilon_k=\frac{4\pi g c L k}{2(4\pi^{2} c^{2} k^{2}-
\bar{\omega}^{2}L^{2}}.
\label{linear}
\end{equation}
Eqs.~(\ref{others}) and (\ref{linear}) give approximate solutions to the 
eigenfrequencies $\Omega_{k},\;\;k=1,2...$.

To solve Eq.~(\ref{eigenfrequencies1}) with respect to the lowest 
eigenfrequency $\Omega_{0}$, let us assume that it satisfies the condition 
$\Omega_{0}L/2c<<1$ (we will see below that this condition is compatible 
with the condition of a small $L$ as defined above). Inserting the 
condition $\Omega_{0}L/2c<<1$ in Eq.~(\ref{eigenfrequencies1}) and keeping 
up to quadratic terms in $\Omega$ we obtain the solution for the lowest 
eigenfrequency  $\Omega_{0}$,
\begin{equation}
\Omega_0=\frac{\bar{\omega}}{\sqrt{1+\frac{\pi gL}{2c}}}.
\label{firsts}
\end{equation}
Consistency between Eq.~(\ref{firsts}) and the condition  
$\Omega_{0}L/2c<<1$ gives a condition on $L$,
\begin{equation}
L\ll 
\frac{2c}{g}f\;;\;\;\;f=\frac{\pi}{2}\left(\frac{g}{\bar{\omega}}\right)^2
\left(1+\sqrt{ 1+\frac{4}{\pi^2}\left(\frac{\bar{\omega}}{g}\right)^2}~
\right).
\label{rsmall}
\end{equation}
Let us consider as in the preceding section, the situations of {\it weak}
coupling, $\kappa^{2}\gg 0$ ($g\ll \bar{\omega}$) and of {\it strong} 
coupling, $\kappa^{2}\ll 0$ ($g\gg \bar{\omega}$). We define a parameter 
$\beta$, by $g=\beta \bar{\omega}$ and weak or strong coupling are defined 
respectively for $\beta\ll 1$ or $\beta \gg 1$. For weak or strong 
couplings we obtain from Eq.~(\ref{rsmall}),
\be
f_{weak}\approx \beta\;;\;\;\;\;\; f_{strong}\approx \frac{\pi}{2}\beta^{2}
\label{betas}
\ee
Let us consider the situation where the dressed material body is initially 
in its first excited level. Then from Eq.~(\ref{eq21}) we obtain the 
probability that it will still be excited after a ellapsed time $t$, 
\begin{eqnarray}
|f^{00}(t)|^{2}&=&(t_{0}^{0})^{4}+2\sum_{k=1}^{\infty}(t_{0}^{0})^{2}
(t_{0}^{k})^{2}\cos(\Omega_{k}-\Omega_{0})t+\nonumber\\
& & ~~~~~~~\sum_{k,l=1}^{\infty}(t_{0}^{k})^{2}
(t_{0}^{l})^{2}\cos(\Omega_{k}-\Omega_{l})t.
\label{|f00R|2}
\end{eqnarray}

{\it a)} {\it Weak} {\it coupling}\\
In the case of {\it weak} coupling a physically interesting situation is 
when interactions of electromagnetic type are involved. In this case, we 
take $\beta=\alpha$, where $\alpha$ is the fine structure constant, 
$\alpha=1/137$. Then the factor $f$ multiplying $2c/g$ in Eq.~(\ref{rsmall}) 
is $\sim 0.07$ and the condition $L\ll 2c/g$ is replaced by a more 
restrictive one, $L\ll 0.07(2c/g)$. For a typical infrared frequency, for 
instance $\bar{\omega}\sim 2,0\times 10^{11}/s$, our calculations are valid
for a value of $L$, $L\ll 10^{-3}m$.

From Eqs.~(\ref{t0r2}) and using the above expressions for the 
eigenfrequencies for small $L$, we obtain the matrix elements,
\begin{equation}
(t_0^0)^2\approx 1-\frac{\pi g L}{2c};\;\;(t_0^k)^2\approx \frac{g L}
{\pi c k^2}.\label{too}
\end{equation}
To obtain the above equations we have neglected the corrective term 
$\epsilon_{k}$, from the expressions for the eigenfrequencies $\Omega_{k}$. 
Nevertheless, corrections in $\epsilon_{k}$ should be included in the 
expressions for the matrix elements $t_{k}^{k}$, in order to avoid spurious 
singularities due to our approximation.
Using Eqs.~(\ref{too}) in Eq.~({\ref{|f00R|2}), we obtain
\begin{eqnarray}
|f^{00}(t)|^{2}&\approx &1-\pi \delta+4(\frac{\delta}{\pi}-\delta^{2})
\sum_{k=1}^{\infty}\frac{1}{k^{2}}\cos(\Omega_{k}-\Omega_{0})t+\nonumber\\
& &~~~\pi^{2}\delta^{2}+\frac{4}{\pi^{2}}\delta^{2}\sum_{k,l=1}^{\infty}
\frac{1}{k^{2}l^{2}}\cos (\Omega_{k}-\Omega_{l})t,
\label{f002}
\end{eqnarray}
where we have introduced the dimensionless parameter $\delta=Lg/2c\;\ll 1$, 
corresponding to a small value of $L$ and we remember that the 
eigenfrequencies are given by Eqs.~(\ref{others}) and (\ref{linear}). As time 
goes on, the probability that the mechanical oscillator be excited attains
periodically a minimum value which has a lower bound given by,
\begin{equation}
\mathrm{Min}(|f^{00}(t)|^{2})=1-\frac{5\pi}{3}\delta+\frac{14\pi^{2}}{9}
\delta^{2}.
\label{min}
\end{equation}
For a frequency $\bar{\omega}$ of the order $\bar{\omega}\sim 4.00\times 
10^{14}/s$ (in the red visible), which corresponds to $\delta\sim 0.005$ and 
$L\sim 1.0\times 10^{-6}m$, we see from Eq.~(\ref{min}) that the probability 
that the material body be at any time excited will never fall below a value
$\sim 0.97$, or a decay probability that is never higher that a value $\sim 
0.03$. It is interesting to compare this result with experimental 
observations in \cite{Haroche3,Hulet}, where stability is found for atoms 
emitting in the visible range placed between two parallel mirrors a distance
$L=1.1\times 10^{-6}m$ apart from one another. For lower frequencies the 
value of the spacing $L$ ensuring quasi-stability of the same order as 
above, for the excited particle may be considerably larger. For instance, 
for $\bar{\omega}$ in a typical microwave value, $\bar{\omega}\sim 2,00
\times 10^{10}/s$ and taking also $\delta \sim 0.005$, the probability that 
the material body remain in the first excited level at any time would be 
larger than a value of the order of $97\%$, for a value of $L$, $L\sim 2.0
\times 10^{-2}m$. The probability that the material body remain excited as 
time goes on, oscillates with time between a maximum and a minimum values 
and never departs significantly from the situation of stability  in the 
excited state.\\

{\it b)} {\it Strong} {\it coupling}\\
In this case we see from Eq.~(\ref{firsts}) that $\Omega_{0}\approx \bar
{\omega}$ for $\beta \gg 2c/\pi L\bar{\omega}$. For $\bar{\omega}\sim 4.00
\times 10^{14}/s$ (in the red visible) and $L\sim 1.0\times 10^{-6}m$ this 
means $\beta \gg 1$. We obtain from Eq.~(\ref{t0r2}),
\begin{equation}
(t_0^0)^2\approx \frac{1}{1+\pi \delta/2};\;\;(t_0^k)^2\approx \frac{g L}
{\pi c k^2}.\label{toos}
\end{equation}
Using Eqs.~(\ref{toos}) in Eq.~({\ref{|f00R|2}), we obtain
\begin{eqnarray}
 |f^{00}(t)|^{2}&\approx &\left(\frac{2}{2+\pi \delta}\right)^{2}+\frac{2}
 {2+\pi \delta}\sum_{k=1}^{\infty}\frac{2\delta}{\pi k^{2}}
 \cos(\Omega_{k}-\Omega_{0})t+\nonumber\\
& &~~~+\frac{4}{\pi^{2}}\delta^{2}\sum_{k,l=1}^{\infty}\frac{1}{k^{2}l^{2}}
\cos (\Omega_{k}-\Omega_{l})t,
\label{f002s}
\end{eqnarray}
The function (\ref{f002s}) As time goes on, the probability that the 
mechanical system be excited attains periodically a minimum value which has 
a lower bound given by,
\be
\mathrm{Min}(|f^{00}(t)|^{2})=\left(\frac{2}{2+\pi \delta}\right)^{2}-
\left(\frac{2}{2+\pi \delta}\right)\frac{\pi \delta}{3}-\frac{\pi^{2}
\delta^{2}}{9}.
\label{mins}
\ee
The condition of positivity of (\ref{mins}) imposes for {\it fixed} values 
of $\beta$ and $\bar{\omega}$ an upper bound for the quantity $\delta$, 
$\delta_{max}$, which corresponds to an upper bound to the diameter $L$ of 
the cavity, $L_{max}$ (remember $\delta=Lg/2c$). Values of $\delta$ larger 
than $\delta_{max}$, or equivalently, values of $L$ larger than $L_{max}$ 
are unphysical and should not be considered. These upper bounds are obtained
from the solution of the inequality $\mathrm{Min}(|f^{00}(t)|^{2})\geq 0$. 
We have $\mathrm{Min}(|f^{00}(t)|^{2})>0$ or 
$\mathrm{Min}(|f^{00}(t)|^{2})=0$, for respectively $\delta<\delta_{max}$ or
$\delta =\delta_{max}$. For a frequency $\bar{\omega}$ of the order 
$\bar{\omega}\sim 4.00\times 10^{14}/s$ (in the red visible), with $\beta=10$ 
($g=10\bar{\omega}$) the lower bound (\ref{mins}) above attains zero for a 
cavity of size $L \sim 1.1\times 10^{-7}m$. For a typical microwave frequency 
$\bar{\omega}\sim 2,00\times 10^{10}/s$, the same vanishing lower bound is 
attained for a cavity of size $L\sim 1.2\times 10^{-3}m$. We see that the 
behaviour of the system for {\it strong} coupling is rather different from 
the weak coupling regime. For appropriate cavity sizes, which are of order 
$10^{-1}$ of those ensuring stability in the weak regime, we ensure for 
strong coupling the complete decay of the system to the ground state in a 
small ellapsed time. in other words, strong coupling {\it enhances} the 
system decay in small cavities, contrarily to the inhibition that happens in
the weak coupling regime.

\section{Concluding Remarks}
We have presented in this paper a non-perturbative treatement of a quantum 
system consisting of particle (in the larger sense of a "material body", an 
atom or a Brownian particle) coupled to an environment modeled by 
non-interacting oscillators. We have used {\it dressed} states which allow 
to divide the system into the {\it dressed} particle and the {\it dressed}
environment by means of some conveniently chosen {\it dressed} coordinates, 
$q_0'$ and $q_j'$ associated respectively to the {\it dressed} particle and 
to the {\it dressed} oscillators composing the environment. In terms of 
these coordinates a division of the system into the {\rm dressed} 
(physically observed) particle and the {\it dressed} environment arises
naturally. The {\rm dressed} particle will contain automatically all the 
effects of the environment on it. This formalism allows a non-perturbative 
approach to the time evolution of a system that may be approximated by a 
particle coupled to its environment, in rather different situations as  
confinement of atoms in cavities or the  Brownian motion. In other words,
underlying  our dressed states formalism there is an unified way to study 
two physically different situations, the radiation process and the Brownian 
motion.  We have approached these situations using the {\it dressed} states, 
and in both we have obtained results in good agreement with experimental 
observations or with expected behaviours. In the Brownian motion we have 
treated in detail the case in which the environment is at zero temperature 
(what corresponds to consider the environment initially in its ground 
state). Our method  accounts for the expected damped behaviour of the 
particle in a much more simpler way than the usual path integral approach. 
For atomic systems we recover with our formalism the experimental 
observation that excited states of atoms in sufficiently small cavities are 
stable. We are able to give formulas for the probability of an atom to 
remain excited for an infinitely long time, provided it is placed in a 
cavity of appropriate size. For an emission frequency in the visible red, 
the size of such cavity is in good agreement with experimental observations 
(\cite{PhysLA}, \cite{physics0111042}). The generalization of the work
presented in this paper to the case of a generic (supraohmic or subohmic) 
environment and finite temperature is in progress and will be presented 
elsewhere.

\vspace{0.5cm}
\begin{center}
{\large \bf Acknowledgements}
\end{center}

A.P.C.M was supported by Conselho Nacional de Desenvolvimento 
Cient\'{\i}fico e Tecnol\'ogico - CNPq and Funda\c{c}\~ao de Amparo 
\`a Pesquisa do Estado do Rio de Janeiro - Faperj
(Brazil). 
G.F.H was supported by a grant from CNPq.

\end{document}